\documentclass[aps,prl,twocolumn,preprintnumbers,superscriptaddress,dblfloatfix,nofootinbib]{revtex4-1}
\usepackage[utf8]{inputenc}
\usepackage{lmodern}
\usepackage{amsmath,amssymb}
\usepackage{graphicx}
\usepackage{url}
\usepackage{color}
\usepackage{enumitem}
\usepackage{subfigure}
\usepackage[dvipsnames]{xcolor}
\usepackage[colorlinks=true,breaklinks=true]{hyperref}
\hypersetup{allcolors=[rgb]{0.0 0.0 0.6},linkcolor=[rgb]{0.75 0.05 0.05}}
\usepackage{bm}
\usepackage{epsfig}
\usepackage{amsmath}
\usepackage{amssymb}
\usepackage{slashed}
\usepackage{color}
\usepackage{accents}
\usepackage[dvipsnames]{xcolor}
\usepackage[colorlinks=true,breaklinks=true]{hyperref}
\hypersetup{allcolors=[rgb]{0.0 0.0 0.6},linkcolor=[rgb]{0.75 0.05 0.05}}

\renewcommand\[{\left[}

\newcommand{\exclude}[1]{}

\begin{document}

\preprint{IPMU20-0092}

\title{Primordial black holes from long-range scalar forces and scalar  radiative cooling}

\author{Marcos M.  Flores} 
\affiliation{Department of Physics and Astronomy, University of California, Los Angeles \\ Los Angeles, California, 90095-1547, USA} 
\author{Alexander Kusenko} 
\affiliation{Department of Physics and Astronomy, University of California, Los Angeles \\ Los Angeles, California, 90095-1547, USA}
\affiliation{Kavli Institute for the Physics and Mathematics of the Universe (WPI), UTIAS \\The University of Tokyo, Kashiwa, Chiba 277-8583, Japan}

\date{\today}

\begin{abstract}
We describe a new scenario for the formation of primordial black holes (PBHs). In the early Universe, the long-range forces mediated by the scalar fields can lead to formation of halos of heavy particles even during the radiation-dominated era. The same interactions result in the emission of scalar radiation from the motion and close encounters of particles in such halos.  Radiative cooling due the scalar radiation allows the halos to collapse to black holes.  We illustrate this scenario on a simple model with fermions interacting via  the Yukawa forces. The abundance and the mass function of PBHs are suitable to account for all dark matter, or for some gravitational wave events detected by LIGO.  The model relates 
the mass of the dark-sector particles to the masses and abundance of dark matter PBHs in a way that can explain why the dark matter and the ordinary matter have similar mass densities. 
The model also predicts a small  contribution to the number of effective light degrees of freedom, which can help reconcile different measurements of the Hubble constant.
\end{abstract}
\maketitle


Primordial black holes (PBHs) formed in the early Universe can account for all or part of dark matter~\cite{Zeldovich:1967,Hawking:1971ei,Carr:1974nx,Khlopov:1985jw,Yokoyama:1995ex,GarciaBellido:1996qt,Kawasaki:1997ju,Green:2004wb,Khlopov:2008qy,Carr:2009jm,Frampton:2010sw,Kawasaki:2016pql,Carr:2016drx,Inomata:2016rbd,Pi:2017gih,Inomata:2017okj,Garcia-Bellido:2017aan,Georg:2017mqk,Inomata:2017vxo,Kocsis:2017yty,Ando:2017veq,Cotner:2016cvr,Cotner:2017tir,Cotner:2018vug,Sasaki:2018dmp,Carr:2018rid,Banik:2018tyb,1939PCPS...35..405H,Cotner:2019ykd,Kusenko:2020pcg,deFreitasPacheco:2020wdg,Takhistov:2020vxs}. Furthermore, PBHs can seed supermassive black holes~\cite{Bean:2002kx,Kawasaki:2012kn,Clesse:2015wea}, can play a role in the synthesis of heavy elements~\cite{Fuller:2017uyd,Takhistov:2017nmt,Takhistov:2017bpt}, and can be responsible for some of the gravitational wave events detected by LIGO~\cite{Abbott:2016blz,Abbott:2016nmj,Abbott:2017vtc,Clesse:2016vqa,Bird:2016dcv,Sasaki:2016jop}.  High energy density in the early Universe facilitates formation of PBHs in the presence of large perturbations from inflation~(e.g.,  \cite{Yokoyama:1995ex,GarciaBellido:1996qt,Kawasaki:1997ju}) or from the scalar field dynamics~\cite{Cotner:2016cvr,Cotner:2017tir,Cotner:2018vug,Cotner:2019ykd}. The scalar forces can generate  instabilities~\cite{Khlopov:1985jw,Kusenko:1997si} leading to  PBHs~\cite{Khlopov:1985jw,Cotner:2016cvr,Cotner:2017tir,Cotner:2018vug,Cotner:2019ykd}.  However, in this class of scenarios, PBHs can only form from rare, overdense, spherical halos, while the rest of the halos virialize and remain mechanically stable until the decay of their constituent particles, $Q$ balls or oscillons~\cite{Cotner:2019ykd}.  Scalar force instability can lead to a growth of structures and formation of halos of interacting particles even during the radiation dominated era~\cite{Gradwohl:1992ue,Gubser:2004uh,Nusser:2004qu,Amendola:2017xhl,Savastano:2019zpr}, and it was conjectured that such early growth of structure could produce  PBHs~\cite{Amendola:2017xhl}; but, unlike the scalar field fragmentation scenarios~\cite{Cotner:2016cvr,Cotner:2017tir,Cotner:2018vug,Cotner:2019ykd}, the growth of structure in the  matter composed of elementary  particles leads to virialized halos, not PBHs~\cite{Gradwohl:1992ue,Gubser:2004uh,Nusser:2004qu,Savastano:2019zpr,Cotner:2019ykd}. 

We describe a new scenario for PBH formation, which is simple and generic: in its minimal realization it involves only one species of heavy particles interacting via the Yukawa forces mediated by a scalar field.  The same long-range scalar interactions that cause the formation of halos during the radiation dominated era~\cite{Gradwohl:1992ue,Farrar:2003uw,Gubser:2004uh,Nusser:2004qu,Amendola:2017xhl,Savastano:2019zpr} allow for emission of scalar waves, which drain energy from the  virialized halos and facilitate their collapse to PBHs.   

Let us consider a fermion $\psi$ interacting with a scalar field $\chi$: 
\begin{equation}
 {\cal L} \supset \frac{1}{2}m_\chi^2\chi^2+ m_\psi\bar{\psi}\psi 
 -y \chi \bar{\psi}\psi +...
 \label{eq:lagrangian}
\end{equation}

We assume that the Universe was radiation dominated at temperatures $T> m_\psi$, and that the $\psi$ particles had equilibrium density.   
We will also assume that the particle number is preserved by an approximate symmetry, and we allow an asymmetry $\eta_\psi= (n_\psi-n_{\bar{\psi}})/s \neq 0$ to  develop, in analogy with the baryon asymmetry $\eta_{\rm B}$, as in the asymmetric dark matter models~\cite{Petraki:2013wwa,Zurek:2013wia}.  
We will assume that the $\chi$ field is either massless or very light, $m_\chi \ll m_\psi^2/M_P$, and that the  $\psi$ particles are either stable or have a total decay width $\Gamma_\psi \ll m_\psi^2/M_P$, where $M_P=M_{\rm Planck}/\sqrt{8\pi}\approx 2\times 10^{18}\ {\rm GeV}$ is the reduced Planck mass, so that there is a  cosmological epoch during which the $\psi$ particles are nonrelativistic, decoupled from equilibrium, and they interact with each other via an attractive long-range force mediated by the $\chi$ field and described by the potential 
\begin{equation}
    V(r)=\frac{y^2}{r} e^{-m_\chi r}.
\end{equation}

During the radiation dominated era, gravitational interactions are not sufficient to allow for a linear growth of structures.  However, scalar forces are usually (and, possibly, always~\cite{ArkaniHamed:2006dz,Palti:2017elp,Gonzalo:2019gjp,Kusenko:2019kcu}) stronger than gravity,  $\beta\equiv y (M_P/m_\psi) \gg 1$, and such forces can cause the fluctuations in the $\psi$ particle number to grow even in the radiation dominated era~\cite{Gradwohl:1992ue,Gubser:2004uh,Nusser:2004qu,Amendola:2017xhl,Savastano:2019zpr}.  We note that the scalar forces couple not to the mass density, but to the $\psi$ number density, and the halos of  $\psi$ particles  grow in the otherwise uniform background of radiation as a form of an isocurvature perturbation. 

The adiabatic density perturbations $\delta(x,t)=\delta \rho/\rho$  grow only logarithmically during the radiation dominated era.  However, the presence of a long-range ``fifth force" stronger than gravity  causes the fluctuations $\Delta (x,t) = \Delta n_\psi/n_\psi $ for an out-of-equilibrium population of heavy, nonrelativistic $\psi$ particles to grow rapidly, as long as $\psi$ is decoupled from radiation, so that the pressure can be neglected.  

For the model of Eq.~(\ref{eq:lagrangian}), if the mean free path of $\chi$ particles in a halo of $\psi$ particles is longer than the size of the halo, the halo is not subject to radiative pressure due to the $\chi $ radiation.  The temperature at which it is true for the  Hubble size halos, and the structures start growing, is $T_g\sim m/[\ln(y^4 M_P/m)]$. This temperature is also close to the temperature $T_f$ at which the annihilation reactions $\bar{\psi} \psi \rightarrow \chi\chi$ freeze-out, which, for $y\sim 1$ and $\eta_\psi \ll 1$ result in the value $T_f\sim m/36$~\cite{Graesser:2011wi}.

In Fourier space, the growth of these perturbations below $T_f$ is described by the system of coupled equations~\cite{Gradwohl:1992ue,Gubser:2004uh,Nusser:2004qu,Amendola:2017xhl,Savastano:2019zpr}

\begin{align}
&\ddot{\delta}_k + \frac{1}{t}\dot{\delta}_k
- \frac{3}{8t^2}(\Omega_r \delta_k + \Omega_{m}\Delta_k)
= 0\label{eq:radEquation}\\[0.25cm]
& \ddot{\Delta}_k + \frac{1}{t}\dot{\Delta}_k
-
\frac{3}{8t^2} [\Omega_r \delta_k + \Omega_{m} (1 + \beta^2)\Delta_k]
= 0,
\label{eq:pStarting}
\end{align}
where $\Omega_r=\rho_r/(\rho_r+\rho_\psi)$ and $\Omega_\psi=\rho_\psi/(\rho_r+\rho_\psi)$  are the radiation and matter fractions, respectively, and $\Omega_r+\Omega_m=1$. Assuming that only radiation and $\psi$ particles are present, and anticipating that all the $\psi $ particles will end up in PBHs, which also scale as matter, the time dependence of these fractions before the matter-radiation equality, $t<t_{\rm eq}$ is given by $\Omega_r=[1+\sqrt{t/t_{\rm eq}}]^{-1}$ and    
$\Omega_m=[1+\sqrt{t_{\rm eq}/t}]^{-1}$.  In the limit $\beta \gg 1$, the perturbations grow fast:
\begin{align}
\Delta_k(t) \approx
\frac{\Delta_k(t_0)}{\sqrt{8\pi}}
\frac{\exp\left(4\sqrt{p}\ (t/t_{\rm eq})^{1/4}\right)}{p^{1/4}(t/t_{\rm eq})^{1/8}},
p = \sqrt{\frac{3}{8}(1 + \beta^2)}.
\end{align}
For $p\gg 1$, the timescale $\tau_\Delta \equiv \Delta_k/(d\Delta_k/dt) $ is  shorter than the Hubble time, which implies a very rapid structure formation.  Thus, in the limit of a strong Yukawa force, the structures form almost instantaneously on all scales up to the horizon size as soon as the $\psi$ particles decouple.  This process was studied in the past, but the fate of the nonlinear structures was not elucidated.  In Ref.~\cite{Amendola:2017xhl}, it was conjectured that the structures could form black holes, but it was later realized  that, instead, these structures remain as virialized dark matter clumps~\cite{Savastano:2019zpr}.   In the absence of energy  dissipation, the latter conclusion is correct because virialization puts an end to any further contraction of halos,  unless energy and angular momentum can be transferred out of the contracting halo. 
 
 
 However, the same long range forces that cause the growth of structure in the $\psi$-particle fluid also cause any particles moving with an acceleration to emit scalar waves, which can dissipate energy from a halo.  This is the key element of PBH formation in the system of matter particles interacting by long-range attractive forces.

 A virialized halo of $N$ particles interacting by scalar Yukawa forces has the potential energy 
$E \sim \frac{y^2 N^2}{R}$, 
where $R$ is the characteristic size of the halo.  Each particle is a source of a scalar field which can be thought of as classical and long range on the length scales shorter than $m_\chi^{-1}$.  A collection of $N$ particles moving inside the halo can radiate scalar waves in several ways.  

First, if the motion is coherent, a dipole moment rotating with a frequency $\omega$ can produce a dipole radiation $P_{\rm coh}\propto y^2 N^2$.  However, for a system of $N$ identical particles, the dipole moment about the center of mass is identically zero because the charge is proportional to the mass, and the first moment of the mass distribution is zero (by the definition of the center of mass).  The coherent quadrupole radiation is possible, but it  is suppressed.  

Second, if each particle is treated as an incoherent source of radiation, the radiated power is proportional to the square of the orbital acceleration $a=\omega^2 R$, where $\omega$ can be different for different particles.  The radiated power $P_{\rm incoh}\propto y^2 \omega^4 R^2 N$ scales as the first power of the number of particles.  This is the correct picture of radiative energy losses in the limit of relatively low number density of particles. 

Third, there is scalar bremsstrahlung  radiation similar to free-free emission of photons from plasma~\cite{Maxon:1967,Maxon:1972}.  Unlike the usual plasma with two charges of particles, our system has identical particles, so the leading bremsstrahlung radiation in two-particle collisions is quadrupole, not dipole, and it is similar to the $e-e$ component of the free-free emission from plasma~\cite{Maxon:1967,Maxon:1972}. 

Finally, if the contracting halo becomes opaque, the radiation is trapped, and the halo turns into a fireball of temperature $T_{\rm halo}$.  
This happens when the collapse timescale $\tau_{\rm coll} = R(t)/(dR/dt)=\sqrt{m/y^2 n_\psi}$ is shorter than the diffusion timescale for $\chi$ radiation $\tau_{\rm diff}\sim 3 R^2/\lambda_\chi$, where $\lambda_\chi$ is the mean free path of the $\chi$ particle in the halo. If, initially, the halo radius is $R_0$ and the density is $n_\psi=\eta_\psi T^3$, the mean free path $\lambda_\chi =1/(\sigma n_\psi)\sim (4\pi m^2/\eta_\psi T^3)(R/R_0)^3 $.  As the size of the halo $R$ decreases, the collapse timescale $\tau_{\rm coll} \sim (m^{1/2}/y T^{3/2} \eta_\psi^{1/2})(R/R_0)^{3/2}$ decreases, while the diffusion timescale $\tau_{\rm diff}\sim (y^4 \eta_\psi T^3 R_0^3)/(4\pi m^2 R)$ increases.  Eventually, the radiation is trapped when diffusion is slower than the collapse, $\tau_{\rm coll} < \tau_{\rm diff}$. 
 
When the $\chi $ radiation is trapped, the cooling proceeds from the surface, and it  can be approximated by the black-body radiation with the power $P_{\rm surf} \sim 4\pi R^2 T_{\rm halo}^4$.  The energy transfer inside the halo can proceed either by diffusion or by convection, and the latter dominates. 
For large $\beta \gg 1$, the scalar force gradients (which exceed the gravitational accelerations) overwhelm the viscosity, leading to very large Rayleigh numbers and fast convection timescales.    The timescale for convective transport is $\tau_{\rm conv}\sim \eta/(R\rho g)$, where $g=y^2N/mR^2$, and $\eta\sim 10 T^3$ is viscosity~\cite{Hosoya:1983xm}, leading to the Rayleigh number ${\rm Ra}  \sim y^2 N R T^2/m\gg 1$, which indicates the halo is highly convective, and convection dominates the heat transport from the core to the surface.

Each of these mechanisms can reduce the energy of the halo on some characteristic timescale.  The energy loss timescale is given by

\begin{equation}
    \tau =\frac{E}{dE/dt} =\frac{E}{P_{\rm incoh}+P_{\rm ff}+P_{\rm surf}...},
\label{eq:tau}
\end{equation}
where 
\begin{align}
E &\sim \frac{y^2 q^2 N^2}{R},
\\ P_{\rm incoh} &\sim \frac{y^6 q^6 N^3}{4m^2R^4},
\label{eq:incoh}\\
 P_{\rm ff}   &\sim \frac{y^6q^6 N^2T_{\rm eff}}{m^2R^3}
 \ln\left(\frac{2T_{\rm eff}}{m}\right)\\
 &\sim 
\frac{y^8q^8 N^3}{m^2 R^4}\ln\left(\frac{Ny^2q^2}{m R}\right) ,
\label{eq:Pff} \\
P_{\rm surf} &\sim 4\pi R^2 T_{\rm halo}^4 =
4\pi \frac{y^2 q^2 N^2}{R^2},
\label{eq:surface}
\end{align}
where $T_{\rm eff}\sim y^2 N/R$ is the energy per particle before the radiation is trapped, while $T_{\rm halo}\sim \sqrt{yN}/R$ is the temperature of trapped radiation after thermalization, as discussed below.  Here $q=1$ for a single $\psi$ particle, while a clump of particles in orbital motion can have $q\gg 1$. The particle mass includes the finite-temperature corrections, $m=m_\psi+(y/4)T$~\cite{Weldon:1982bn}.

When the particle density is very low, the incoherent emission (\ref{eq:incoh}) is the dominant channel for the energy loss.  However, when the mean separation between particles is smaller than the radiation length, the radiation from the neighboring particles can interfere, and Eq.(\ref{eq:incoh}) is not applicable.  
However, since the structure we consider exists on a broad range of scales, small clumps rotating in the larger halo can radiate as ``particles" in Eq.(\ref{eq:incoh}) with $q\gg 1$.  In the absence of $N$-body simulations, we cannot reliably count on this dissipation channel.  Therefore, we will base the discussion on the bremsstrahlung emission (\ref{eq:Pff}), yielding a conservative estimate, which can only be helped by any additional dissipation. 

A halo of size $R$ can lose energy and contract to a black hole at temperature $T$ as long as $\tau(R) < M_P/T^2 $. Since the timescale is an increasing function of the halo size, the halos with smaller $R$, for which $\tau < M_P/T^2 $, collapse first.  Those halos for which $\tau (R) > M_P/T^2 $ may never collapse if the formation of PBH from smaller halos eliminates the long-range scalar forces. 

Initially, the halo of size $R$ has a potential energy $\sim y^2N^2/R_i$, and it initially radiates with the power $P_{\rm ff}$, Eq. (\ref{eq:Pff}).  As the halo contracts, the number density increases and the $\chi$ radiation is trapped forming a fireball of temperature $T_{\rm halo}$ that can be estimated from energy conservation: $-y^2N^2/R_i=
 -y^2N^2/R(t)+(4\pi/3)R(t)^3 T_{\rm halo}^4 $.  This implies the halo temperature $T_{\rm halo }\sim \sqrt{y N}/R(t) $. The solution for the size of the halo determined by $dE/dt=P_{\rm surf} $, which implies $R(t) =  R(0)(1 - t/\tau_{\rm surf})$. 
As the halo starts to shrink, the characteristic timescale $\tau$ decreases, leading to even faster energy dissipation.  This signals collapse of the halo to a PBH.  

At high densities, the $\psi$ particles can form bound states with discrete quantum levels, and the emission picture changes to that which is similar to hot gas emitting photons.  The viscosity and the ram pressure of such a gas of ``atoms"  can speed up the process of collapse into a black hole. 

This very simplified thermal history involves two stages: the initial cooling by bremsstrahlung, until the radiation is trapped, and the following cooling from the surface of a hot fireball.  The bremsstrahlung timescale $\tau_{\rm ff}$ is the longer of the two, and it serves as the bottleneck limiting the collapse  of the largest halos.  

We find that for a wide range of parameters and $y\gtrsim 10^{-3}$, the radiative cooling timescale in either the high-density, low-density, or intermediate-density regimes is smaller than the Hubble time.  Therefore, the collapse of a halo to a black hole is possible and it proceeds unimpeded as the radius decreases and reaches the Schwarzschild radius.  

Formation of black holes halts further structure evolution because, in accordance with the no-hair theorems, black holes do not carry global charges and do not feel the long-range forces due to scalar interactions of particles that fell into the black holes. The strong  long-range forces are likely to cause all or most of the $\psi$ particles to end up in PBH.  The cosmological PBH abundance is then equal to the $\psi$ particle abundance, and their fraction at present time is related to the baryon density:

\begin{equation}
f_{\rm PBH}=\frac{\Omega_{\rm PBH}}{\Omega_{\rm DM}}=
0.2 \frac{m_\psi}{m_p} 
\frac{\eta_\psi}{\eta_{\rm B}}= \left ( \frac{m_\psi}{5 \, {\rm GeV}} \right ) \left ( \frac{\eta_\psi}{10^{-10}} 
\right ) .
\label{eq:fpbh}
\end{equation}
Therefore,  our scenario has the same potential to explain the closeness of $\Omega_{\rm DM}$ and $ \Omega_{\rm B}$, as the models with asymmetric particle dark matter~\cite{Petraki:2013wwa,Zurek:2013wia}.  The asymmetry $\eta_\psi$ can arise from the same process that produces the baryon asymmetry of the universe.

Let us now estimate the mass function of PBHs starting with the smallest masses. The limit $N> (M_P/m)^2$ can be derived by requiring that, as $R$ approaches the Schwarzschild radius $R_S=mN/M_P^2$, the halo is still larger than the Compton wavelength of the $\psi$ particle.  It is unlikely that a black hole would form from a halo with fewer particles than $N_{\rm min}=(M_P/m)^2$. 
For fermions $\psi$, one also needs to require that, as the Fermi degeneracy is reached in the course of a  collapsing halo, the Fermi energy be small compared to the potential energy $y^2 N/R$ as $R\rightarrow R_S$.  This condition turns our to be less constraining than the quantum condition $N>N_{\rm min}$. We note that the Chandrasekhar limit of $N> (M_P/m)^3$ derived for the gravitational potential is effectively weakened here by a factor $(m/y M_P)^2\ll 1 $. A naive lower limit on the mass of a halo that can form a PBH could be set as $M>m N_{\rm min}=M_P^2/m=5\times 10^{-21} M_\odot (1~{\rm GeV}/m)$. However, it is unlikely that a black hole could form close to the quantum uncertainty  limit. Viscous friction, tidal friction, and gravitational mergers cause multiple neighboring halos to merge and form a single black hole, hence increasing the minimal size.  We parametrize the minimal PBH mass in the form
\begin{align}
     M_{\rm min} &= \zeta N_{\rm min} m
     = 10^{-15} M_\odot 
    \left (\frac{\zeta}{10^6}\right )   
    \left ( \frac{5\, {\rm GeV}}{m}
    \right ).
    \label{eq:Mmin}
\end{align}
Here $\zeta = F_{\rm visc} F_{\rm mergers}$, where $F_{\rm visc} $ is the effect of viscous friction and tidal effects that could lead to merger of neighboring dense halos into one, and $F_{\rm mergers}$ represents the effects of gravitational merger of black holes.  The exact values of these factors require detailed analysis and numerical simulations. We assume that the viscosity and the gravitational tidal forces act at least on the length scales of the order of $~(10-100) R$, in a volume that encompasses more than $10^3$ halos, so that $F_{\rm visc} \gtrsim 10^3$,  $F_{\rm mergers}\gtrsim 10^3 $, leading to $\zeta \sim 10^6$, which we will use as an illustrative value.  

Since the PBH formation is rapid and takes about one Hubble time, the mass function of PBHs should represent a snapshot of the structure in the $\psi$ fluid at the time of formation. In the absence of $N$-body simulations, the details of the $\psi$ halo structure formation are not known, but the structure can be described approximately.  Since the collapsing halos are formed from the growth of perturbations followed by a short history of mergers, the resulting PBH mass function can be approximated by a Press-Schechter function: 
 \begin{equation}
    M \frac{dN_h}{dM}\propto\frac{1}{\sqrt{\pi}} \left (\frac{M}{M_*} \right)^{1/2}e^{-M/M_*}.
 \end{equation}

\begin{figure}[htb]
\includegraphics[width=0.95\linewidth]{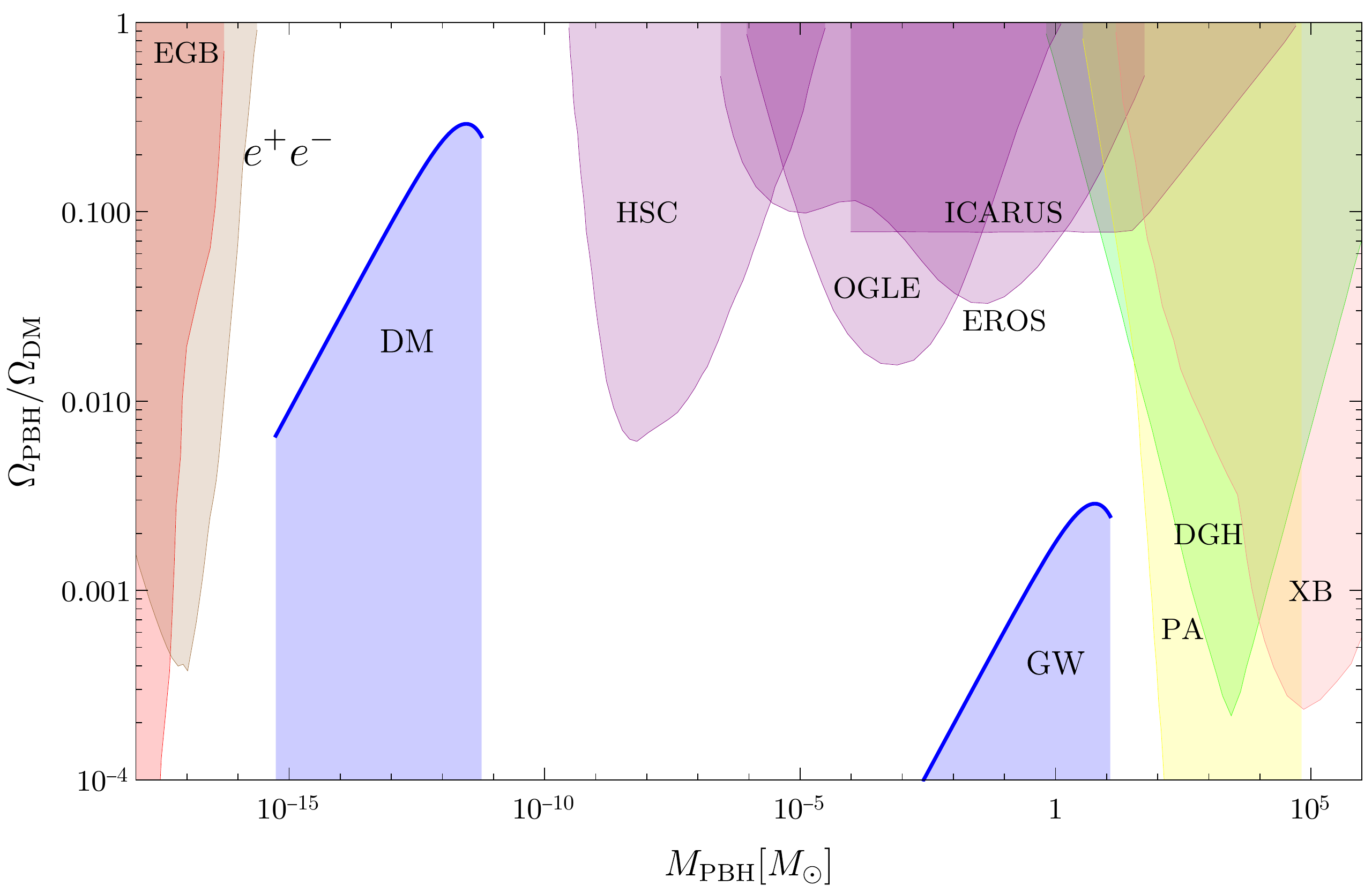}
\caption{The mass functions of PBH (line labeled  ``DM") can account for all dark matter if the asymmetry in the dark sector is the same as the baryon asymmetry, $\eta_\psi \sim \eta_{\rm B}\sim 10^{-10}$, in a model with $m_\psi= 5$~GeV and $y=5\times 10^{-3}$.   The PBHs are in the mass range of interest to LIGO, Virgo, and KAGRA  (line labeled ``GW") for $m_\psi=5$~MeV, $y=1.5\times 10^{-2}$, $\eta_\psi = 10^{-9}$.  
The constraints are from Refs.~\cite{Ali-Haimoud:2016mbv,Niikura:2017zjd,Niikura:2019kqi,Inoue:2017csr,Carr:2020xqk,Carr:2020gox,Lu:2020bmd, Dasgupta:2019cae, Laha:2019ssq}. }
\label{fig:massfunction}
\end{figure}

The characteristic mass $M_*$ is set by the largest size $R_*$ for which the emission timescale $\tau(R_*)$ in Eq.(\ref{eq:tau}) is smaller than the Hubble time. For the relevant range of parameters, the main emission channels are bremsstrahlung ($\tau\sim \tau_{\rm ff}$) followed by the radiative cooling from the surface ($\tau\sim \tau_{\rm surf }$).  Since $\tau_{\rm ff}>\tau_{\rm surf}$, it is the bremsstrahlung timescale $\tau_{\rm ff}$ that determines whether or not a given halo has time to collapse before the smaller halos become black holes and terminate the action of the long-range forces.  Solving for the size $\tau_{\rm ff} (R_*)=t_H$, we find the characteristic mass 
\begin{equation}
M_*= \frac{4\pi}{3}m n_\psi R_*^3, \ \ 
R_* \simeq 
3\times 10^4\times 
\left(
\frac{\eta_\psi M_{P}^4 y^6}{g_*m^7}
\right)^{1/3}
.
\end{equation}
We can parametrize $M_*$ in the form
\begin{align}
M_* &\simeq 2\times 10^8\ \frac{\eta_\psi^2M_{P}^4 y^6}{m^3}\\
&\simeq
6\times 10^{-12}\ {\rm M}_\odot
\left(
\frac{\eta_\psi}{10^{-10}}
\right)^{2}
\left(
\frac{5\ {\rm GeV}}{m_\psi}
\right)^{3}\times\\
& \times \left(
\frac{y}{5\times 10^{-3}}
\right)^{6}
.
\end{align}
The resulting mass function is shown in Fig.~\ref{fig:massfunction} for our model with $m_\psi=5$~GeV, $\eta_\psi=10^{-10}$.  

The $\chi$ particle mass $m_\chi$ must be small enough to allow for the long-range forces.  If $m_\chi > T_f^2/M_P$, the long-range force cuts off at distances $R\sim 1/m_\chi$, resulting in the upper limit on the size of the characteristic scale in the Press-Schechter function,  $R_*<1/m_\chi$. 

The radiative cooling of a collapsing halo is a complex dynamical problem.  We have neglected the spatial density and temperature distributions and the existence of smaller halos inside larger halos, as well as screening of the long-range forces by the finite density and temperature corrections to the scalar mass~\cite{Ayaita:2012xm,Casas:2016duf}, which in turn depend on the density distribution.  These effects can be studied in numerical $N$-body simulations. 
If the collapse is delayed by some dynamics not captured by our discussion, the delay allows larger structures to form and collapse, extending the mass function toward larger masses.  

Our scenario can be realized in a variety of models with different degrees of complexity in the dark sector.  The simplest  model described by the Lagrangian (\ref{eq:lagrangian}) is particularly appealing.   Let us assume that the asymmetry in the dark sector is similar to the baryon asymmetry of the Universe as in  popular models of asymmetric dark matter~\cite{Petraki:2013wwa,Zurek:2013wia}.  Then the abundance of PBH (\ref{eq:fpbh}) is just right to explain all dark matter for $m_\psi = 5$~GeV. 
The resulting mass function of PBHs, shown in Fig.~\ref{fig:massfunction} by a solid line labeled ``DM", is consistent with all present observations and can account for all dark matter.  

We note that, if $m_\psi\gg 5$~GeV, the black holes are small, and they can evaporate before the structure formation.  So, a more complex dark sector involving multiple heavy particles could still result in PBH dark matter dominated by the contribution of the $\sim 5$~GeV species.  
This strengthens the naturalness argument: if the asymmetry in the dark sector is the same as the baryon asymmetry, and if there is a tower of dark states with different masses, the $\sim 5$~GeV mass produces all the dark matter, and any contribution of the heavier particles is naturally eliminated, because the resulting PBHs are small enough to evaporate. 

This model predicts a slightly different value for the effective number of degrees of freedom than the standard $N_{\rm eff}=3.05$.  If one assumes that the dark sector, comprising $\psi$ and $\chi$ particles, had the same temperature as the visible sector at $T\sim m_\psi$, one can estimate the contribution $\Delta N_{\rm eff}$ of the light $\chi$ particles to radiation. In the dark sector, the number of effective light degrees of freedom goes from $g_1=1+(7/8)\times 2$ to $g_2=1$. This  contributes to the measured value of  $N_{\rm eff}$~\cite{Blennow:2012de,Fuller:2011qy,Patwardhan:2015kga}:
\begin{equation}
    \Delta N_{\rm eff} = 14 [g_1/g_*(T_d)]^{4/3}\approx 0.1-0.2,
\end{equation}
where the model-dependent temperature for decoupling between the visible and the dark sectors is taken to be in the range $T_d = 1-100$~GeV. 
The value $\Delta N_{\rm eff}\sim 0.2$ is allowed, and, in fact, it was argued that $\Delta N_{\rm eff}=0.3-0.4$ can reconcile the local and the  cosmological measurements of the Hubble constant~\cite{Bernal:2016gxb,Gelmini:2019deq,Anchordoqui:2019yzc,Vattis:2019efj,Escudero:2019gvw,Gelmini:2020ekg,Vagnozzi:2019ezj,Wong:2019kwg}.  

Another interesting set of parameters leads to the mass function of interest to gravitational waves detectors shown in Fig.~\ref{fig:massfunction} and labeled ``GW".  For $m_\psi=5$~MeV, $y=1.5\times 10^{-2}$, $\eta_\psi=10^{-9}$, the resulting mass function extends to $M_*\gtrsim 10 M_\odot$, with a sufficient abundance to explain some of the  events reported by LIGO~\cite{Abbott:2016blz}. 

Our scenario leads to PBH clustering that resembles the fully formed nonlinear structure at the time of their formation.  This, as well as departure from spherical symmetry in the collapse of each halo imply that the gravitational waves background and the distribution of spins can be very different from those expected from other PBH formation mechanisms. 

In summary, we have presented a novel scenario for the formation of primordial black holes.  The  scalar fields that mediate long-range attractive forces enable both the  clustering of heavy particles and the radiative cooling by emission of scalar waves.  The cooling facilitates  collapse of the halos into black holes, which can account for all dark matter.  In the example using decoupled fermions interacting by the Yukawa forces, the resulting PBH dark matter density is related to the particle mass and can naturally explain the dark matter abundance. 

\begin{acknowledgments}
We thank K.~Petraki, J.~Rubio, M.~Sasaki, V.~Takhistov, and E.~Vitagliano for helpful discussions.  
This work was supported by the U.S. Department of Energy (DOE) Grant No. DE-SC0009937.  A.K. was also supported by the World Premier International Research Center Initiative (WPI), MEXT, Japan and by Japan Society for the Promotion of Science (JSPS) KAKENHI Grant No. JP20326504.
\end{acknowledgments}

 
\bibliography{bibliography}
 
\end{document}